\documentclass[12pt]{article}

\usepackage{etex} 

\usepackage[margin=3cm]{geometry}
\usepackage{amsmath}
\usepackage{amssymb}
\usepackage{amsthm}
\usepackage{fancyhdr}
\usepackage{verbatim}
\usepackage{mathtools}
\usepackage{color}
\usepackage[utf8]{inputenc}

\numberwithin{equation}{section}

\theoremstyle{plain}

\theoremstyle{definition}

\begin{document} 

\title{Typical ground states for large sets of interactions} 
\author{Aernout van Enter \\ Bernoulli Institute, Nijenborgh 9 \\ Groningen University, \\9747AG, Groningen, Netherlands\\
a.c.d.van.enter@rug.nl
\\ \\ Jacek Mi\c{e}kisz \\ Institute of Applied Mathematics and Mechanics \\ University of Warsaw \\ Banacha 2, 02-097 Warsaw, Poland \\ miekisz@mimuw.edu.pl} 
\pagenumbering{arabic} 

\baselineskip=20pt
\maketitle 

\noindent
{\it Dedicated to Joel Lebowitz on the occasion of his 90th birthday.\\ With best wishes and many thanks for being such a stimulus and such an  inspiration,\\  as a researcher, as an editor, as a conference organiser,  and  as a human being.\\  He has been a role model for both of us for  as many years as we have  been working in science.}

\begin{abstract}
We discuss what ground states for generic interactions look like. We note that a recent result, due to Morris, implies that the behaviour of ground-state measures 
for generic interactions is similar to that of generic measures. In particular, it follows from his observation that they have singular spectrum and that they are weak mixing, but not mixing.
\end{abstract}


\section{Introduction - generic behaviour}
It is an important question of statistical physics what the properties of ground states, Gibbs states, phase diagrams are.

 It is common to start this question in the context of classical statistical mechanics, and often the problem is even more  simplified by studying  the zero-temperature question, the ground state problem. Until around 1980 there was a consensus among condensed-matter physicists that it was to be expected that usually ground states, and in higher dimensions also low-temperature equilibrium states, are crystalline, displaying some kind of periodicity.

One justification for this  belief was the fact that the densest packing of hard spheres was known (in $d=1$ or $d=2$, now also in $d=3$, $d=8$ and $d=24$ \cite{Hale, Via, CKMRV}) or conjectured (in various other dimensions)  to be periodic. It is possible to rephrase this densest-packing question as the study of  the behaviour of the ground states for a classical gas with hard-core interactions. 

Densest packings of more general and varying shapes, however,  have led to the study of  tiling problems. It was discovered, by Robinson and later by Penrose, that there exist finite sets of tiles which enforce quasiperiodicity. Again, tiling problems can be rephrased as ground state problems for nearest-neighbour interactions,  where different tiles represent different particles, and even before the experimental discovery of quasicrystals, in mathematics the assumption of ubiquitous periodic behaviour was starting to be put in doubt. For an early description of the connection between tiling problems and ground state problems see e.g. \cite{Rad3, Rad4} or \cite{MieLeu}.

Another research field developed by the study of ground states for  particular, physically plausible, models. In various examples, in particular  for Lennard-Jones and similar potentials,  mostly in d=1,  see e.g. \cite{GR}, but also more recently for d=2. see e.g. \cite{DF, Theil}, it was proven that ground states are periodic.   

The Crystal Problem has been reviewed in \cite{Rad3}, later in \cite{LL} and even more recently in \cite{BL}.

One further simplification we will employ (as has been also discussed to some extent in the above reviews)  is that we  study lattice models. This considerably simplifies the problem, as any underlying periodic structure, which in continuous models is very hard to show, even though it might occur  (whether due to some hard-core or to some Lennard-Jones-like terms in the interaction), now comes for free. On top of that, an aperiodic placing of particles on lattice sites could occur. 
 
 Moreover one might argue,  especially in lower dimension at positive temperatures, that lattice models may not be that physically  realistic. Still, the study of lattice models has proved to be invaluable in the the understanding of various forms of long-range order occurring in a wide variety of physical systems. And the more serious objections to using lattice models, even at $T=0$,  do not apply with the same force to understanding what happens in  longer-range models, nor do they really  apply  to attempts  to understand what may occur in the physical dimension $d=3$.  

One question this issue of various types of long-range order  has led to, and which we will discuss here, is about the behaviour of equilibrium or ground states  for typical interactions. An old result due to Gallavotti and Miracle \cite{GalMir}, using a theorem of Mazur \cite{Maz}, implies that the set of interactions having a unique tangent to the pressure (= equilibrium state), is {\em generic} (that is, it is a dense $G_{\delta}$, a countable intersection of dense open sets). This result holds in any of the commonly studied (Banach) spaces of lattice interactions.
The largest interaction space usually considered,  $B_0$, can be associated to the space of continuous functions, modulo translations, see e.g. \cite{Isr1}. In this space, a number of ``pathological'' properties can be proven, e.g. there is a dense set of interactions having uncountably many ergodic equilibrium or ground states, \cite{Isr1,Isr3,Shino, ST}, the pressure is never Fr\'echet differentiable \cite{DE}, the pressure can depend on boundary conditions, and ``Ideally Metastable'' states  \cite{Sew} as well as "frozen" low-temperature states \cite{Fis,BruiLep1, BruiLep2} can exist.

Later results were proven on the triviality of generic phase diagrams (implying the generic violation of the Gibbs Phase Rule)  for various spaces of long-range interactions \cite{Isr2,IsrPh} . 

More recently, the study of ground states in the guise of ``ergodic optimisation'' \cite{Boch,Jenk} has reinvigorated the interest in such questions. 
For the space of continuous functions, it was proven that  ground states for generic interactions have full support, and zero entropy \cite{BoJenk, Bre,YZ}. 

Even more recently, Morris \cite{Mor} proved a result that properties of generic measures are inherited by ground state measures of generic continuous functions. He applied this to show that they are not mixing.\\ 
Here we add the observation in our {\bf Corollary 2} that they are weak mixing and have singular (Dynamical and thus Diffraction) spectrum. This follows from the known equivalent statements for generic measures, due to Halmos, Rohlin, and Knill \cite{Hal,Roh,Kn}.

Another, maybe not too surprising, consequence, using another  result of Israel \cite{Isr4}, is that ground states for typical  (generic) interactions cannot be Gibbs states (for possibly different interactions).  

We remark, by the way, that Simon's "Wonderland" theorem has before  provided a number of other examples of problems where singular spectrum of some kind turns out to be generic \cite{Sim1}.  

Moreover we discuss and speculate on properties of ground states and equilibrium  states (which then can be Gibbs states in the DLR sense) in smaller interaction spaces. 

One of the fundamental problems in statistical physics is to understand why matter at low temperatures and high enough pressures possesses some sort of long-range order. For ages this  was interpreted to mean that matter is crystalline, that is, its constituents, atoms or molecules, form some kind  of a three-dimensional lattice. The famous, and still not solved, Crystal Problem is to show that for "reasonable" physical interactions between particles, the arrangements minimizing their energy density are attained by periodic configurations. Periodicity is the strongest embodiment of a positional long-range order. 

One major physical reason which has spurred the interest in the behaviour of "typical" ground states was the discovery of quasicrystals \cite{shechtman}. Despite earlier beliefs and claims that all or most physical systems should have crystalline, periodically ordered, ground states and low-temperature states, such claims  have turned out to be dubious and in many contexts untrue. For an early rigorous result contradicting this crystalline paradigm, see for example \cite{MieRad}. Especially since Shechtman's discovery of the first quasicrystals \cite{shechtman}, which led to his 2011 Nobel Prize in Chemistry, there has developed a large amount of research, experimental, theoretical and also mathematical,  about the properties of  quasicrystals, the nature of the associated aperiodic order and related questions. See for example the books  \cite{BaaGri, Sene}.

For a recent, more physics-style paper, again showing that periodic order is generically not to be expected,  see for example \cite{FrSt}. 

The discovery of quasicrystals showed us that other forms of long-range order than periodic ones might be present in Nature. The non-periodic order of quasicrystals was represented by a (dense) discrete spectrum in X-ray experiments. From a different point of view, energy-minimizing (or free-energy-minimizing) configurations of particles gave rise to non-mixing ground-state (or Gibbs state) measures. Many (toy) examples were constructed with such properties, some of them based on previously constructed non-periodic tilings \cite{Rad0,jm1,jm2,jm3,jm4} The natural question then arises: how typical (generic) and how robust are such examples? It was proven in 
\cite{Mie,MieRad,Rad1} that in the Banach space of two-body summable interactions, for typical interactions, that is in a dense $G_{\delta}$ set, the ground-state measure is non-periodic, non-mixing and has a zero entropy. 
In \cite{jm3}, a classical lattice-gas model was constructed with a non-periodic ground-state measure which is stable against small perturbations of nearest-neighbour interactions.

In this note we review and combine some old results on generic properties of lattice-gas models with recent ideas from ergodic optimization, and present the new result that for generic interactions a ground state is non-mixing, but weakly mixing, and moreover it has a singular (dynamical and thus diffraction) spectrum. It means that generically ground states are quite disordered but will still have some long-range order. 

Our result is based on the recent result of Morris \cite{Mor} on ergodic optimization for generic continuous functions. In particular he showed that ergodic measures which maximize the integral of a generic continuous {\em function} have the same properties as generic ergodic {\em measures}.

In Section 2, we introduce classical lattice-gas models.
 We also review some old results concerning generic presence of quasi-crystalline or weak crystalline ground states.

 In Section 3, we show how our result follows from that of Morris.
 
 In Section 4, we discuss some open problems and directions of future research.    

\section{Classical lattice-gas models}
We will consider the case of classical, finite-spin lattice systems, as for example discussed in \cite{EFS,FriVel,Geo,Isr1,Rue1,Sim}.
Our formulation mainly follows Israel \cite{Isr1}.

Our configuration space is $\Omega= {\Omega_0}^{Z^d}$, with $\Omega_0$ finite.
Translations on $Z^d$ are indicated by $\tau_x, x \in Z^d$.

Translation-invariant interactions $\Phi$ are sets of functions  $\Phi_X$ on 
${\Omega_0}^X$, with $X$ a finite subset of $Z^d$, and such that the $\Phi_X$ are translation-invariant.

Different interaction (Banach) spaces can be defined by different norms, using  different translation-invariant functions $f$ on the subsets of $Z^d$, by 
$|||\Phi_X|||_f = \sum_{0 \in X} ||\Phi_X|| f(X)$. \\Often the $f$ one chooses depends on either $|X|$, the cardinality of $X$, or on its diameter $diam(X)$. 
The largest interaction space we will consider is $B_0$, which is obtained by choosing $f(X)= \frac{1}{X}$.  Other commonly used interaction spaces are $B_n$,  defined by taking $f(X)= {|X|}^{n-1}$, and $B_{\lambda,exp}$, defined by $f(X) =e^{ \lambda |X|}$. \\
If  it is the case that $\Phi_X =0$ for all $X$ with large enough diameter, we say that our interaction is of finite range; if it is the case that  $\Phi_X =0$ when $X$ contains more than two sites, we say  that we are considering  pair interactions. 

To each interaction $\Phi$ is associated a continuous function on $\Omega$, describing the energy per site, localised around the origin,   
$A_{\Phi} = \sum_{0 \in X}\frac{\Phi_X}{|X|}$. Translation-invariant measures on $\Omega$
correspond to bounded linear functionals on $B_0$, in an isometric way (see e.g. \cite{Isr1}, Lemma II.1.1).

On $B_0$ one can define a Lipschitz continuous and convex pressure function $P$. On the set of translation-invariant probability measures on $\Omega$ one can define an affine entropy  (density)  function $s$.

Pressure and entropy are each other's Legendre-Fenchel transforms, and are related by dual variational principles:

\begin{equation}
P(\Phi) = sup ( s(\mu) - \mu(A_{\Phi})| \mu \in E_I)
 \end{equation}
and 
\begin{equation}
s(\mu) = inf (P(\Phi) + \mu(A_{\Phi}| \Phi \in B_0)  
\end{equation}  
 
Solutions of these variational principles satisfy
\begin{equation}
P(\Phi) = s(\mu_{\Phi}) - \mu_{\Phi}(A_{\Phi}).
\end{equation}
In the case that a measure  $\mu_{\Phi}$ solves the variational principle, we say that $\mu_{\Phi}$ is an equilibrium state for $\Phi$. Such an equilibrium state corresponds to a tangent functional to the pressure function, tangent to P at the point $\Phi$. 

If we replace the affine entropy function $s(\mu)$ by $0$, the corresponding measure becomes a ground state measure. 

{\bf Remark:} In ergodic optimization, see e.g. \cite{Boch,Jenk}, usually one considers maximizing, rather than minimizing (ground state) measures, but the questions are easily seen to be equivalent by a simple sign change.
 
The fact that convex functions on Banach spaces generically have a unique tangent implies that generically there exists a unique translation invariant equilibrium or ground state \cite{GalMir,Rue2}. 


\section{Generic measures and generic interactions, corollaries of a theorem by Morris}

In \cite{Mor} the following Theorem was proven. In our setting it says the following:

\noindent
{\bf Morris' Theorem:}\\
Let  $U$ be a generic set of translation-invariant measures on $\Omega$. Then the set of functions $V$  whose ground states are in $U$ is generic  (a $G_{\delta}$ in the space of continuous functions on $\Omega$).

\smallskip

In words, it says that if a generic set of measures has a certain property, this same property holds for the ground states of a generic set of interactions in $C(\Omega)$ (or in $B_0$).

\noindent
{\bf Corollary 1 (Morris):}\\ In particular, it was concluded  by Morris   that generic ground states (= maximizing measures) are unique, non-mixing, have full support, and have entropy zero.

\noindent
{\bf Corollary 2:}\\
 Ground states for generic interactions in $C(\Omega)$ are  weak mixing and have singular diffraction spectrum. Moreover they cannot be written as Gibbs measures for any interaction. 
 
\smallskip 

\noindent
{\bf Proof:} \\
This follows directly from the fact that those properties are known to hold for generic, translation-invariant measures. These results are due respectively to Halmos \cite{Hal}, Knill \cite{Kn}, and Israel \cite{Isr4}. 

\smallskip

We notice that in $C(\Omega_0)$ there are dense sets of interactions with uncountably many ergodic ground states \cite{Shino, Isr3}. Thus one cannot expect much more regular behaviour. We do remark, though, that the absence of point spectrum in the pure singular spectrum result shows that generic ground states are neither periodic, nor quasiperiodic. Thus generically neither crystals, not quasicrystals exist (but ``weak'' \cite{EntMie} or ``turbulent''\cite{Rue3}  crystals do). Thus is compatible  with the results of \cite{Mie,MieRad,Rad1,Rad2}, showing some statistical homogeneity, but excluding periodicity, for generic interactions in various interaction spaces. 
 
\section{ Speculations on different interaction spaces}

The space $B_0$, although mathematically natural, as one can associate it to the continuous functions, has a number of pathological properties. There exist dense sets (although not generic ones) for which there are uncountably many ergodic (extremal translation-invariant) equilibrium or ground states \cite{Isr3,Shino}, -as is proven via the Bishop-Phelps theorem- generic ground states have full support \cite{BoJenk}, the pressure can depend on boundary conditions, strict convexity of the pressure does not hold, see e.g. the discussion in Section 2.6.7 of  \cite{ EFS}.

This is a reason why often smaller interaction spaces are considered. In the space $B_1$, Gibbs measures can be defined, according to the prescriptions of Dobrushin, Lanford and Ruelle, in $B_2$, the uniqueness theorem of Dobrushin implies that there are open high-temperature, high-magnetic-field or low-density regimes. In $B_{\lambda,exp}$ it can be shown that pressure and states can be analytic in open high-temperature or low-density regions, etc.

Also one can have open  sets in those smaller spaces  where there are pure, homogeneous ground state configurations (vacua). Typical statements about ground states then distinguish between two cases, the case where there is a unique homogeneous ground state configuration and the case where there is some weak, non-periodic form of long-range order. We suspect that the lack of point spectrum, that is the lack of ``crystalline'' long-range order,  may also hold there.  

We  remark that Israel \cite{Isr2} has proved that phase coexistence is exceptional, in these interaction spaces. His arguments are written down for positive temperature, but also apply for ground states. It seems that his arguments with minor modifications can be used to obtain a genericity statement excluding periodicity in ground states, in a somewhat more abstract and general context, than the result excluding periodicity for generic interactions as was obtained for lattice-gas models in \cite{MieRad}. In other words, under a generic long-range perturbation of an interaction a unique pure phase (or vacuum) can be stable, but neither phase coexistence, nor periodicity are stable under such perturbations.  

One question we don't know how to answer, however, is what happens with aperiodic order under long-range perturbations. Quasicrystalline long-range order is defined in terms of discrete diffraction spectra, whereas weaker forms of long-range order occur when singular spectra appear (weak or turbulent crystals \cite{EntMie,Rue3}), or even can occur without any spectral indications \cite{Slaw}. As the number of possible periods is countable, excluding all of them leads to considering a  countable intersection of dense $G_{\delta}$ sets, which preserves genericity; the number of possible quasicrystalline spectra is uncountable, so a different argument would be required to investigate stability questions of quasicrystalline order. 

On the other hand, if we restrict ourselves even more and look at fast decaying interactions in one dimension, corresponding for example with
 H\"older or Lipschitz functions, sometimes it can be proven that typical ground states behave quite differently, and that periodic ground states are generic \cite{Cont, HLMXZ}.

We note that it has been known for quite some time that any finite-range interaction in one dimension always has periodic ground states \cite{ban,RadSch,MieRad0}. But this is no longer true if one allows even fast decaying interactions in one dimension, or nearest-neighbour interactions based on tilings in higher dimensions, see e.g. \cite{Aub,EKM, GERM,Rad3,Rad4}. 
For example,  it is not known if it is the case that the quasicrystalline order such as occurs in Devil's Staircases \cite{Aub1,Aub2,Aub3,Bak,jjm1,jjm2}, or in more general Sturmian (balanced) ground states \cite{EKM}, is stable under some class of short-range perturbations.
 
In higher dimensions for short-range interactions, however, in many cases long-range order is stable, as can be shown by Pirogov-Sinai theory. However, it is not known if genericity statements in the sense of statements holding true for "generic  short-range interactions", and predicting that they behave in a certain way as regards their long-range order, are valid. 

{\bf Acknowledgements:}  We would like to thank the National Science Centre (Poland) for a financial support under Grant No. 2016/22/M/ST1/00536, which made a visit of AvE to Warsaw possible, during which we started this project.

\end{document}